\documentclass[12pt]{article}

\usepackage{amsfonts}
\usepackage{amsmath}
\usepackage{amssymb}
\usepackage{cite}
\usepackage{comment}

\newcommand{\ba}{\begin{eqnarray}}
\newcommand{\ea}{\end{eqnarray}}
\newcommand{\nn}{\nonumber}
\newcommand{\nt}{\nonumber\\}
\newcommand{\back}{\!\!\!\!\!\!}
\newcommand{\udl}{\underline}

\newcommand{\Winf}{\mathcal{W}_{1+\infty}}
\newcommand{\cN}{\mathcal{N}}
\newcommand{\cW}{\mathcal{W}}
\newcommand{\p}{\partial}
\newcommand{\fU}{\varphi^{U}}
\newcommand{\fV}{\varphi^{V}}
\newcommand{\bD}{{\bar \Delta}}
\newcommand{\bZ}{\mathbb{Z}}

\newcommand{\eps}{\epsilon}
\newcommand{\emp}{\emptyset}

\newcommand{\sstar}{{}^{\raisebox{0.3ex}[0ex][0ex]{${}_\star$}}_{\raisebox{0.6ex}[0ex][0ex]{${}_\star$}}}

\begin{document}
\begin{titlepage}

\begin{flushright}
UT-11-12\\
KEK-TH-1457
\end{flushright}

\vskip 12mm

\begin{center}

{\bfseries\fontsize{19pt}{20pt}\selectfont
$\Winf$ algebra as a symmetry behind AGT relation}

\vskip 12mm
{\large
Shoichi Kanno$^\dagger$\footnote{
E-mail address: kanno@hep-th.phys.s.u-tokyo.ac.jp},
Yutaka Matsuo$^\dagger$\footnote{
E-mail address: matsuo@phys.s.u-tokyo.ac.jp}
and
Shotaro Shiba$^\ddagger$\footnote{
E-mail address: sshiba@post.kek.jp}
 }\\
\vskip 5mm
{\it
$^\dagger$
Department of Physics, Faculty of Science, University of Tokyo,\\
Hongo 7-3-1, Bunkyo-ku, Tokyo 113-0033, Japan\\
}
\vskip 4mm
{\it
$^\ddagger$
Institute of Particle and Nuclear Studies,\\
High Energy Accelerator Research Organization (KEK),\\
Oho 1-1, Tsukuba-city, Ibaraki 305-0801, Japan\\
\noindent{ \smallskip }\\
}
\vspace{45pt}
\end{center}
\begin{abstract}
We give some evidences which imply that $\Winf$ algebra
describes the symmetry behind AGT(-W) conjecture:
a correspondence between the partition function of $\cN=2$ supersymmetric
quiver gauge theories and the correlators of Liouville (Toda) field theory.
\end{abstract}

\end{titlepage}

\setcounter{footnote}{0}

\section{Introduction}

The purpose of this paper is to give a proposal
on the symmetry behind the correspondence between
4-dim $\cN=2$ supersymmetric quiver gauge theory and 2-dim conformal
Liouville (Toda) system, which was discovered by seminal papers by
Seiberg-Witten \cite{Seiberg:1994rs,Seiberg:1994aj}
and later has been deepened by recent breakthroughs 
\cite{Gaiotto:2009we,Alday:2009aq}.  

In particular 
in \cite{Alday:2009aq},  an explicit relation between the two ---
Nekrasov's partition function \cite{b:Nekrasov} 
for $SU(2)$ quiver gauge theories and 
the conformal block of Liouville theory \cite{b:Liouville} ---
was given.  It was then generalized to the $SU(N)$ case, 
where the Liouville theory is replaced by $A_{N-1}$ Toda theory
\cite{Wyllard:2009hg}.

Such correspondence is interesting since
(1) It implies a nontrivial relation between 2D and 4D physics
which may be explained by the strong coupling physics of M-theory 5-brane.
(2) It apparently relates the formulae with very different mathematical origin,
{\em i.e.}~one in the geometry of instanton moduli space of the gauge theories
and the other in the representation theory of  infinite dimensional Lie algebra
such as Virasoro algebra or $W_N$ algebra.

At this stage, this correspondence, called AGT relation 
or AGT-W relation (for its generalization), 
have two issues in different levels
to be fully explored.  First, one needs to know precise
definition of the statement:
\ba\label{AGT}
Z^{\mathrm{Nekrasov}} \!&=&\! \langle V_1\cdots V_n\rangle
\ea
where the left hand side is the partition function of super Yang-Mills theory
and the right hand side is the chiral correlator of Liouville (Toda) field theory.
While the left hand side is well-known for linear quiver
gauge theories \cite{b:Nekrasov}, 
the corresponding chiral correlation function of
Liouville (Toda) theory is only known in the special case  
\cite{Fateev:2005gs}. Furthermore, 
there remain some open issues for the choice of vertex operators 
and intermediate states. 

Second, one needs to understand 
more profound issues why such correspondence exists.
This is certainly much more important for the future development
but it is out of reach of this paper since we would like to
focus on the symmetry of 2D CFT.

Recently, a major step to understand the
right hand side of eq.\,(\ref{AGT}) was undertaken
\cite{Alba:2010qc,Mironov:2010pi,Belavin:2011js}.
The key issue here is how to understand the factorized form of
$Z^{\mathrm{Nekrasov}}$ from CFT viewpoint.
For $SU(2)$ quiver, the authors of \cite{Alba:2010qc} proposed 
a basis with two Young diagram indices $|Y_1, Y_2\rangle$ 
where the factors in Nekrasov's partition function are reproduced 
through the norm and 3-point functions
in terms of them.  When one of $Y_i$'s is null, the basis coincides with
Jack symmetric polynomial\cite{b:Stanley}.  
They also presented an algorithm to construct such basis for general cases.
Later, for the 
simpler case (where the central charge $c=1$ or $Q=0$), 
it was conjectured in \cite{Belavin:2011js}
that the state $|Y_1, Y_2\rangle$ is given by the direct product of 
Schur polynomials $|Y_1, Y_2\rangle=s_{Y_1} s_{Y_2}$.

Other than these technical improvements toward the proof of AGT-W relation, 
these studies reveal the importance of
integration of somewhat mysterious 
``$U(1)$ factor"  \cite{Alday:2009aq} to construct 
these useful bases. This implies that the original
symmetry such as Virasoro or $W$ algebra should be properly enhanced to
include the $U(1)$ factor.  In this paper, we propose that
$\Winf$ algebra, whose representation was studied long ago
\cite{Kac:1993zg,b:AFMOQ,Frenkel:1994em,Awata:1994tf}, 
should be the proper symmetry behind AGT-W relation, at least for $Q=0$.

As its strange name implies, $\Winf$ algebra contains
$U(1)$ current operator together with the infinite number of
higher spin generators.  With appropriate choice of representation,
we will show that the algebra reduces to $W_N$ algebra and $U(1)$ current 
as expected from above constructions.
It is known further that Schur polynomial gives
an appropriate basis which diagonalizes all the commuting charges. 
Therefore, it is very natural that this $U(1)$ current corresponds to the 
$U(1)$ factor,
and this discussion gives some indirect evidences why $\Winf$ algebra
is relevant to AGT-W relation.

This paper is organized as follows. 
In \S\,\ref{s:review}, we briefly review the definition of 
$\Winf$ algebra and its representation. 
In particular, we emphasize the importance of ``quasi-finite"
representation \cite{Kac:1993zg}.  This section is a brief 
summary of \cite{Awata:1994tf}.
In \S\,\ref{s:reduction}, we demonstrate
explicitly how the representation of $\Winf$ algebra with the central charge
$C=N$ reduces to that of $W_N$ algebra together with $U(1)$ current.
In \S\,\ref{s:conjecture}, we discuss that AGT-W relation 
is reduced to the problem of computation of 3-point function
of $\Winf$ algebra.  After presenting our conjecture,
we show some evidences by generalizing the computation
of \cite{Belavin:2011js} to $W_3$ cases.
In \cite{Belavin:2011js}, the proof of AGT-W conjecture
was reduced to so called ``chain vector".
In \S\,\ref{sec:chain}, we review what a chain vector is
and the derive their explicit form for Virasoro and $W_3$
case.  Compared with \cite{Belavin:2011js}, our novelty is
the use of free boson variable from the beginning (this helps to
simplify the computation) and derivation of $W_3$ chain vector.
After this preparation, in \S\,\ref{s:comb}, we combine
the $U(1)$ factor as predicted by $\Winf$ algebra and reproduces
Nekrasov's formula.
In \S\,\ref{s:conclusion}, we illustrate the future directions.
In the appendix, we give a summary of our notation.

\section{A brief review of $\Winf$ algebra and its representation}
\label{s:review}

In the following, we briefly review some relevant material of
the representation theory of $\Winf$ algebra. We follow the description
of \cite{Awata:1994tf}.

\subsection{$\Winf$ algebra}

$\Winf$ algebra is a quantum realization of
algebra generated by higher order differential operators
$z^m D^n$ ($D:=z\frac\partial{\partial z}$, $m\in \bZ$, $n=0,1,2,\cdots$). 
We define a map it to quantum operator through
$z^m D^n \rightarrow W(z^m D^n)$.
$\Winf$ algebra is most compactly expressed in the following form:
\ba
[W(z^n e^{xD}, W(z^m e^{yD})] \!&=&\! (e^{mx}-e^{ny})
W(z^{n+m} e^{(x+y)D}) -C\frac{e^{mx}-e^{ny}}{e^{x+y}-1}\delta_{n+m,0}
\ea
where $C$ is the central extension parameter and 
$W(z^n e^{xD}):=\sum_{m=0}^\infty\frac{x^m}{m!} W(z^n D^m)$.

The algebra contains $U(1)$ current operators $J_m=W(z^m)$.
There are some ambiguity in the choice of Virasoro operators.
One may take, for example,  $-W(z^n D)$
which satisfies Virasoro algebra with central charge $-2C$.  
For this choice, however,
the U(1) currents have anomalous transformation
$
\left[J_n, W(z^m D)\right]=-n J_{n+m}+\frac{C}2 n(n-1)\delta_{n+m,0}\,.
$
Then a better choice is
\ba
L_n=-W(z^n D)-\frac{n+1}{2} W(z^n)
\ea
with which $J_n$ transforms as the primary field with spin 1.
This operator satisfies the Virasoro algebra with central charge $C$.  
Together with these familiar ones, $\Winf$ algebra also contains infinite
 number of higher spin operators $W(z^n D^m)$ whose commutation relation
 with Virasoro operator is
\ba
[L_n, W(z^m D^l)]=(ln-m)W(z^{m+n} D^l)+\cdots\,.
\ea
The first term implies that these operators transform as spin $l+1$ fields,
but the algebra contains extra terms $\cdots$ which implies that they should
be modified to be primary fields.  
We will come back to this problem for spin 3 case in \S\,\ref{s:WN}. 

The algebra has an infinite number of commuting charges
$W(D^n)$ ($n=0,1,2,\cdots$) and we need their eigenvalues
to specify the representation.  As usual,
the highest weight state (HWS) $|\Delta\rangle$ is defined by
\ba
W(z^n D^m)|\Delta\rangle \!&=&\! 0 \qquad\quad\,\, (n>0,~m\geq 0)\nn\\
W(D^n)|\Delta\rangle \!&=&\! \Delta_n |\Delta\rangle \quad (n\geq 0)
\ea
where $\Delta_n$ ($n=0,1,\cdots$) are complex number parameters to specify
the representation.
They are more conveniently expressed in a form of generating function
\ba
W(e^{xD})|\Delta\rangle=-\Delta(x) |\Delta\rangle\,,
\quad
\Delta(x):=-\sum_{n=0}^\infty \frac{x^n}{n!}\Delta_n \,.
\ea
The Hilbert space is generated from HWS by applying
$W(z^{-n} D^m)$ ($n>0$). 
Since $[L_0, W(z^{-n}D^m)]=nW(z^{-n}D^m)$, 
the inner products of such states are block diagonal with respect to 
the eigenvalue of $L_0$ which we call ``level" of the state. 

\subsection{Quasi-finite representation}

Unlike the usual 2D chiral algebra, $\Winf$ algebra contains
infinite number of states $W(z^{-n}D^m)|\Delta\rangle$ ($m=0,1,2,\cdots$)
at each level $n$. 
It makes the handling of Hilbert space quite difficult.
 
To make the situation better, we require 
a condition on $\Delta(x)$ (first discovered by \cite{Kac:1993zg})
such that most of the states at each level except for 
a finite set become null.  Such
representation is called ``quasi-finite representation".
It is realized  by requiring conditions of the form
\ba
W(z^{-n} b_n(D))|\Delta\rangle\sim 0
\ea
where $b_n(x)$ is a polynomial of $x$.  If such condition is imposed, 
all operators of the form $W(z^{-n} D^m b_n(D))|\Delta\rangle$ become null, 
thus there remain only finite number of
operators $W(z^{-n} D^l)$ with $l<\mbox{order}\,(b_n(x))$ 
for each level $n$.  The polynomials $b_n(x)$ are determined from
that for level 1 $b_1(x):=b(x)$ through the consistency with the algebra:
\ba\label{bn}
b_n(x)=\mbox{lcm}\,(b(x),\,b(x-1),\,\cdots,\,b(x-n+1))
\ea
where `lcm' means the least common multiple, and 
$b(x)$ is called as the characteristic polynomial.
In order to have such null states, $\Delta(x)$ needs to satisfy
\ba\label{cDE}
b\left(\frac{d}{dx}\right) ((e^x-1)\Delta(x)+C)=0\,.
\ea
In particular, for $b(x)=\prod_{i=1}^K (x-\lambda_i)^{m_i}$ 
with $\lambda_i\neq \lambda_j$, the solution to eq.\,(\ref{cDE}) is
\ba\label{sDE}
\Delta(x)=\frac{\sum_{i=1}^Kp_i(x) e^{\lambda_i x} -C}{e^x-1}
\ea
where $p_i(x)$ is a polynomial of degree $m_i-1$
and satisfies $\sum_i p_i(0)=C$.

We note here that 
$\Winf$ algebra has
a one-parameter family of automorphism which is 
called ``spectral flow".  The transformation rule is
\ba
\tilde W(z^n e^{xD})=W(z^n e^{x(D+\lambda)})-C\frac{e^{\lambda x}-1}{e^x-1}
\delta_{n0}
\ea
where $\tilde W$ satisfies the same algebra as $W$,
but their eigenvalues for $|\Delta\rangle$ are modified.
For the representation (\ref{sDE}), this transformation 
is realized as a shift
$\lambda_i\rightarrow \lambda_i+\lambda$.
It implies that the representation obtained by
shift of $\lambda_i$ by $\lambda$ has exactly the same
property as the original one.

\paragraph{Unitary representations}
In order to make the Hilbert space unitary, we need to impose further
constraints on $\Delta(x)$ \cite{b:AFMOQ, Frenkel:1994em}.
It may be summarized as follows.

Firstly, the multiplicity indices $m_i$ in $b(x)$ should be one.
Then the solution (\ref{sDE}) for $m_i=1$ becomes
\ba\label{sDE2}
\Delta(x)=\sum_{i=1}^K C_i \frac{ e^{\lambda_i x} -1}{e^x-1}\,,\quad
\sum_{i=1}^K C_i=C\,.
\ea
Secondly, the parameters $C_i$ in eq.\,(\ref{sDE2}) must be positive integer.
In particular, for $C_i=1$ (for all $i$) 
and $\lambda_i-\lambda_j\neq$ integer (for all pair $i\neq j$),
we have a free fermion representation
\ba
&&\back
b^{(i)}(z) =\sum_{r\in \bZ} b^{(i)}_r z^{-r-\lambda_i-1},\quad
c^{(i)}(z)  =\sum_{r\in \bZ} c^{(i)}_r z^{-r+\lambda_i},\quad
b^{(i)}(z)c^{(j)}(w)\sim \frac{\delta_{ij}}{z-w}\,,\nt
&&\back 
b^{(i)}_r|\Delta \rangle =c^{(i)}_s|\Delta\rangle =0\quad
(r\geq 0,~ s\geq 1)\,,\quad
c^{(i)\dagger}_s=b^{(i)}_{-s}\,,
\ea
and
\ba
&&\back
W(z^n e^{xD}) = \sum_{i=1}^K\left(\sum_{r+s=n} e^{x(\lambda_i-s)} E^{(i)}(r,s) 
-\frac{e^{\lambda_i x}-1}{e^x-1} \delta_{n,0}
\right),\quad
E^{(i)}(r,s) = \sstar b^{(i)}_r c^{(i)}_s \sstar\,,
\qquad\label{e:W}
\ea
where the normal ordering $\sstar\,\,\,\,\sstar$ is defined as
$\sstar b^{(i)}_r c^{(i)}_s \sstar=b^{(i)}_r c^{(i)}_s $ if $r\leq -1$
and $-c^{(i)}_s b^{(i)}_r$ if $r\geq 0$.
Note that the parameter $K$ in eq.\,(\ref{sDE2}) equals to $C$ in this case.
The Hilbert space is the tensor product of free fermions
with fermion number $=0$ for each $i$.  
A convenient basis of such states is labeled by $K$ Young diagrams
$\vec Y=(Y_1,\cdots, Y_K)$. 
For example, for $K=1$ case, 
the state associated with $Y=[f_1,\ldots,f_r]$ ({\em i.e.}~length
of rows are $f_1\geq\cdots\geq f_r\geq 1$) is given by\footnote{
For other representations, see for example, appendix B of \cite{Awata:1994tf}.}
\ba
|Y\rangle = b_{-\bar f_1} b_{-\bar f_2}\cdots b_{-\bar f_r} |-r\rangle
\,,\quad
\bar f_i=f_i-i-1\,,\quad
|-r\rangle = c_{-r+1}\cdots c_{-1} c_0|\Delta\rangle\,.
\ea
The basis $|\vec Y\rangle$ for general $K$ is a tensor product of such states.
After bosonization, such basis is written as the product of Schur polynomials,
as we will see later.  We note that good characterization of such states
is that they are diagonal with respect to $W(D^n)$ action 
as shown in section 3.1 of \cite{Awata:1994tf}.

We note that if some of $\lambda_i$'s satisfy $\lambda_i-\lambda_j=$ integer,
the free fermion basis does not give the Hilbert space of $\Winf$ algebra.
To see this, we should remember the definition of the polynomial $b_n(x)$.
For $\lambda_i-\lambda_j\neq$ integer, $b_n(x)=\prod_{i=1}^n b(x-i)$.
If $\lambda_i-\lambda_j=$ integer, however,
the order of polynomial $b_n(x)$ becomes lower, 
since we have lcm in eq.\,(\ref{bn}).  
It implies that we have extra null states.
Thus the Hilbert space of $\Winf$ becomes in general smaller than
those spanned by free fermions.

\section{Reduction of $\Winf$ to $W_N$ algebra and $U(1)$ factor}
\label{s:reduction}

\subsection{Explicit form of some unitary representations}
Before we start, we explain the structure of representations
with $C=1,2,\cdots$
to some detail.

\paragraph{\udl{$C=1$}}
In this case, $\Delta(x)=\dfrac{e^{\lambda x}-1}{e^x-1}$\,.
By using the spectral flow, one may shift $\lambda\to 0$,
which means $\Delta(x)\to 0$.  In this sense, we have only one
highest weight state. The Hilbert space of $\Winf$ algebra
coincides with that of one free fermion pair with fermion number
zero. Therefore, the partition function becomes 
\ba
Z(q) =\sum_{\mathcal{H}} q^{L_0}= q^{\lambda^2/2}
\prod_{n=1}^\infty \frac{1}{1-q^n}\,.
\ea
\paragraph{\udl{$C=2$}}
In this case, the generating function of the weights $\Delta_n$ becomes
\ba
\Delta(x) =
\frac{e^{\lambda_1 x}-1}{e^x-1}+\frac{e^{\lambda_2 x}-1}{e^x-1}
=\sum_{i=1,2}\left(\lambda_i+\frac12 (\lambda_i^2-\lambda_i)
 x +\frac1{12}(2\lambda_i-1)(\lambda_i-1)x^2+\cdots\right)
\ea
which implies
\ba
J_0|\Delta\rangle = -(\lambda_1+\lambda_2)|\Delta\rangle\,,\quad
L_0|\Delta\rangle=\frac12(\lambda_1^2+\lambda_2^2)|\Delta\rangle\,,\quad
\cdots\,.
\ea
The spectral flow may be used to set $J_0$ eigenvalue to be zero, and so
one may put $\lambda_1=\lambda/2$, $\lambda_2=-\lambda/2$.
Then the conformal weight of $|\Delta\rangle$ becomes $\lambda^2/4$,
which looks like the Virasoro conformal weight for the vertex operator
of a free boson $e^{\pm\lambda \phi/\sqrt{2}}$.
This will be confirmed in the next subsection.

If $\lambda\notin \bZ$,
the partition function is that for two free bosons
\ba
Z=\prod_{n=1}^\infty \frac1{(1-q^n)^2}\,.
\ea
For $\lambda\in \bZ$, however,
the Hilbert space is in general smaller than that of fermionic representation,
since $b_n(x)\neq \prod_{i=1}^n b(x-i)$ for $n>|\lambda|$. 
In particular for $\lambda=0$, $\Delta(x)=0$ with the characteristic polynomial $b(x)=x$.
This implies that $W(z^{-1}D^m)|\Delta\rangle =0$ for $m=1,2,\cdots$.
More explicitly, $W(z^{-1})|\Delta\rangle =\sum_{i=1,2}E^{(i)}(-1,0)|\Delta\rangle$
and 
$W(z^{-1}D)|\Delta\rangle = \sum_{i=1,2}\lambda_i E^{(i)}(-1,0)|\Delta\rangle$.
They are not independent if $\lambda_1=\lambda_2$, 
since the second state is a linear function of the first one.
Therefore, the partition function for $\lambda=0$ becomes
\ba
Z=\frac{1}{1-q}\prod_{n=2}^\infty \frac1{(1-q^n)^2}\,.
\ea

\paragraph{\udl{$C>2$}} Up to spectral flow symmetry, the representation
contains $C-1\,(=N-1)$ parameters. The vertex operator of $W_N$ algebra
has the same number of independent parameters. In fact, we can identify them
as we see in next subsection.

\subsection{Reduction from $\Winf$ to $W_N$}
\label{s:WN}
In order to see the connection with AGT-W relation, 
we need to see the explicit relation with $\cW_N$ algebra.
To see it, we start from the free fermion realization (\ref{e:W})
with $C=K=N$. We introduce free fermion fields
\ba
b^{(i)}(z) = \sum_n b^{(i)}_n z^{-n-1}\,,\quad 
c^{(i)}(z)=\sum_n c^{(i)}_n z^{-n}\,.
\ea
We note that after replacing $c_n=\psi_{n-1/2}$ and $b_n=\bar\psi_{n+1/2}$,
this definition agrees with the standard Dirac fermion $\psi(z)$, $\bar\psi(z)$
in NS sector.
We define a generating function of $W(z^ne^{xD})$ as
\ba
W(\zeta,x)
\!&:=&\! \sum_n W(z^n e^{xD} )\zeta^{-n-1}\nn\\
&=&\! \sum_{i=1}^N \sum_{r,s}\left(
e^{{x}(\lambda_i-s)} \,\sstar b_r^{(i)} \zeta^{-r-1} c_s^{(i)} \zeta^{-s}\sstar
-\frac{e^{\lambda_i x} -1}{e^{x}-1}\zeta^{-1} 
\right)\nn\\
&=&\! \sum_{i=1}^N e^{x\lambda_i} \,\sstar b^{(i)}(\zeta) e^{xD_\zeta} c^{(i)}(\zeta)\sstar -\zeta^{-1}\Delta(x)\nn\\
&=&\! \sum_{i=1}^N e^{x\lambda_i} \,\sstar b^{(i)}(\zeta) c^{(i)}(e^x\zeta)\sstar -\zeta^{-1} \Delta(x)
\ea
where $D_\zeta:=\zeta\partial_\zeta$.
We apply the standard bosonization rule to the fermions
\ba
b^{(i)}(\zeta) = \,:\! e^{-\phi^{(i)}(\zeta)} \!:\,,\quad
c^{(i)}(\zeta) = \,:\! e^{\phi^{(i)}(\zeta)} \!:\,,
\ea
where $:\,\,\,\,\,:$ refers to the normal ordering of bosonic oscillator
and
\ba
\phi^{(i)}(\zeta)= x^{(i)} +\alpha_0^{(i)} \log\zeta 
-\sum_{n}\frac{\alpha^{(i)}_n}{n} \zeta^{-n},\quad
[\alpha^{(i)}_n,\alpha^{(j)}_m]=n\delta_{n+m}\delta^{ij}\,.
\ea
The fermionic normal ordering means 
\ba
\sstar b(\zeta) c(e^x \zeta)\sstar = b(\zeta)c(e^x \zeta)-\frac{1}{\zeta(e^x-1)}\,,
\ea
then the generating function $W(\zeta,x)$ can be written in a simplified form as
\ba
W(\zeta,x)
\!&=&\! \sum_{i=1}^N e^{x\lambda_i}
\frac{-1}{\zeta(e^x-1)}\left(
:\! e^{\phi^{(i)}(e^x\zeta)-\phi^{(i)}(\zeta)}\!: -1
\right)-\frac{1}{\zeta}\Delta(x)\nn\\
&=&\! -\sum_{i=1}^N 
\frac{1}{\zeta(e^x-1)}\left(
:\! e^{\phi^{(i)}(e^x\zeta)-\phi^{(i)}(\zeta)+x\lambda_i}\!: -1
\right).
\ea
Here the exponent function can be written as 
\ba
\phi^{(i)}(e^x\zeta)-\phi^{(i)}(\zeta)+x\lambda_i=(\alpha^{(i)}_0 +\lambda_i) x
+\mbox{(oscillator part)}\,,
\ea
so we can see that 
$\lambda_i$ plays a role of shifting momentum $p^{(i)}=\alpha_0^{(i)}$
in free boson. Therefore, one may rewrite the vertex operator part as
\ba
:\! e^{\phi^{(i)}(e^x\zeta)-\phi^{(i)}(\zeta)+x\lambda_i }\!:\,
= \,:\! e^{\varphi^{(i)}(e^x\zeta)-\varphi^{(i)}(\zeta)+x\lambda_i}\!:
  \,:\! e^{\varphi^{U}(e^x\zeta)-\varphi^U(\zeta)}\!: 
\ea
where 
$\varphi^U(\zeta):=\frac{1}{N}\sum_i \phi^{(i)}(\zeta)$ and
$\varphi^{(i)}(\zeta)=\phi^{(i)}(\zeta)-\varphi^U(\zeta)$.
By using this expression, we can separate $W(\zeta,x)$ into $W_N$ and $U(1)$ part:
\ba
W(\zeta,x)\!&=&\!-\frac{1}{\zeta(e^x-1)}\left(\Xi(\zeta,x)\, \Xi^U(\zeta,x)-N
\right),\label{Wsep}\\
\Xi(\zeta,x)\!&:=&\!\sum_{i=1}^N
:\! e^{\varphi^{(i)}(e^x\zeta)-\varphi^{(i)}(\zeta)}\!:\,,
\quad \Xi^U(\zeta,x)~:=\,\,\,
:\! e^{\varphi^{U}(e^x\zeta)-\varphi^{U}(\zeta)}\!: .
\ea
Here the factor $x\lambda_i$ is absorbed into the redefinition of zero mode
of $\varphi^{(i)}$ field. Note that eq.\,(\ref{Wsep}) tells us
how to decompose the operator into U(1) factor and $W_N$
generators.

Since $\Xi(\zeta)$ is invariant under Weyl reflection of $\varphi^{(i)}$,
we conjecture that the module generated by such 
operators should be rewritten
in terms of $W_n^{(m)}$ ($m=2,3,\cdots,N$).
Let us confirm it for $N=2,3$.
We expand $W(\zeta,x)$ as
\ba\label{expand-W}
W(\zeta,x) \!&=&\! \sum_{n}\zeta^{-n-1} W(z^n e^{xD})\nn\\
\!&=&\! \sum_n \sum_{m=0}^\infty \frac{x^m}{m!}\zeta^{-n-1}W(z^n D^m)\nn\\
\!&=&\! J(\zeta) -\zeta x \bigl(T(\zeta) -\frac12\partial_\zeta J(\zeta)\bigr)
+\frac12 \zeta^2 x^2 \tilde W_3(\zeta)+\cdots
\ea
where $J(\zeta)=\sum_n J_n \zeta^{-n-1}$,
$T(\zeta)=\sum_{n} L_n \zeta^{-n-2}$ and 
$\tilde W_3(\zeta)=\sum_n W(z^nD^2) \zeta^{-n-3}$.
Now we write explicit form of operators in terms of free bosons.

For $N=2$, we write $\varphi^{(1)}=\frac12 (\phi^{(1)}-\phi^{(2)})=:\varphi^V$
and $\varphi^{(2)}=-\varphi^V$. Then eq.\,(\ref{Wsep}) gives
\ba
J(\zeta) \!&=&\! -2\partial\varphi^U \nt
T(\zeta) \!&=&\! T^V(\zeta) +(\partial\varphi^U)^2 \nt
\tilde W_3(\zeta) \!&=&\!
-2(\partial\varphi^U)T^V -\partial T^V -\frac{1}{\zeta}T^V 
\nt
&&\!
-\frac{2}{3} (\partial\varphi^U)^3
 -2 (\partial^2\varphi^U)(\partial\varphi^U)
 -\frac{2}{3}(\partial^3\varphi^U)
 -\frac{1}{\zeta}\bigl((\partial^2\varphi^U)+(\partial\varphi^U)^2\bigr).
\ea
where $T^V(\zeta):=(\partial\varphi^V)^2$. 
The $U(1)$ current operator and Virasoro operator take the standard form
with $C=2$.  Expression for $\tilde W_3$ is complicated, but
the dependence on $\varphi^V$ can be written in terms of only 
$T^V$ and its derivative.  In this sense, we can expect that the Hilbert
space of  $\Winf$ algebra can be expressed in terms of the reduced
set, {\em i.e.}~Virasoro operator $T^V$ and $U(1)$ current $J=-2\partial\varphi^U$.

For $N=3$, we write
\ba\label{n3-varphi}
\varphi^{(1)} \!&=&\! 
 \frac13(2\phi^{(1)}-\phi^{(2)}-\phi^{(3)})=:-\frac{1}{\sqrt{6}}(\varphi^V_1+\sqrt{3}\varphi^V_2)\nt
\varphi^{(2)} \!&=&\!
 \frac13(-\phi^{(1)}+2\phi^{(2)}-\phi^{(3)})=:
-\frac{1}{\sqrt{6}}(\varphi^V_1-\sqrt{3}\varphi^V_2)
\nt
\varphi^{(3)} \!&=&\! 
 \frac13(-\phi^{(1)}-\phi^{(2)}+2\phi^{(3)})
 =\sqrt{\frac{2}{3}}\varphi^V_1\,.
\ea
Then the expression for generators becomes
\ba
J(\zeta) \!&=&\! -3\partial\varphi^U\nt
T(\zeta) \!&=&\! T^V(\zeta)+\frac32 (\partial\varphi^U)^2\nt
\tilde W_3(\zeta) \!&=&\! -\sqrt{\frac{2}{3}} W_3^V
 -(\p\fU)T^V-\frac12\partial T^V-\frac{1}{2\zeta}T^V\nn\\
&&\! -(\p\fU)^3-3(\p \fU)(\p^2\fU)-\p^3\fU-\frac{3}{2\zeta} ((\p\fU)^2+\p^2 \fU)
\ea
where
\ba
T^V(\zeta) := \frac12(\partial \varphi^V_1)^2
+\frac12(\partial \varphi^V_2)^2\,,\quad
W_3^V(\zeta) := \frac16\bigl(
(\p\fV_2)^3-3 (\p\fV_1)^2\p\fV_2\bigr).
\ea
The $U(1)$ current and Virasoro generator are again the standard one for $C=3$.  
In the expression of  $\tilde W_3(\zeta)$, 
$W_3^V(\zeta)$ is a spin 3 primary field with respect to $T^V(\zeta)$
and coincides with the $W_3$ generator for the central charge $c=2$.
The other terms are written in terms of $T^V(\zeta)$, $\p\fU$ and their derivatives.
We conjecture that the higher terms in eq.\,(\ref{expand-W})
can be also written in terms of only $W_3(\zeta)$, $T^V(\zeta)$, $\p\fU$ and their derivatives.
If it is true, the Hilbert space for $C=3$ system is described
by the $W_3$ operators and $U(1)$ part.

\section{Conjecture and some evidences}
\label{s:conjecture}

\subsection{General strategy}
We hope that we have convinced the readers who followed 
\S\,\ref{s:review} and \ref{s:reduction} of the following fact:
$\Winf$ algebra contains the infinite number of 2D chiral fields with
spin $1,2,\cdots,\infty$.  When we limit ourselves to the unitary 
quasi-finite representations,  the central charge $C$ must be a finite positive
integer $N$ and the 
independent chiral fields must be limited to those with spin $1,2,\cdots, N$.  
Among these fields, those with spin $2,3,\cdots, N$ coincide with 
the chiral fields of $W_N$ algebra.  We have shown it
explicitly for $N=2,3$ in \S\,\ref{s:WN},
but its generalization for $N>3$ would be clear through our arguments.
A novelty here is that we also have $U(1)$ current $J(\zeta)$. 
While we may decouple it  from $W_N$  generators in the Hilbert space, 
we need it to realize the larger symmetry $\Winf$.

We may compare the situation in AGT-W relation \cite{Alday:2009aq,Wyllard:2009hg}.  
In these works, the chiral symmetry in 2D side 
is described by $W_N$ algebra with central charge 
$c=N-1+Q^2N(N^2-1)$ in order to be compared with $SU(N)$ quiver gauge theories.
Here the parameter $Q=b+1/b$ corresponds to a set of deformation parameters $\eps_{1,2}$ appearing in Nekrasov's partition function as $Q=\frac{\eps_1+\eps_2}{\sqrt{\eps_1\eps_2}}$.
In order to compare the correlation function of Liouville (more generally, Toda) field theory with Nekrasov's partition function, 
we need extra ``$U(1)$ factor" for the former function \cite{Alday:2009aq}.  

In $\Winf$ approach with the quasi-finite unitary representation, 
we need to restrict ourselves to $Q=0$ and $C=(N-1)+1$, where the former $N-1$ part is described by $W_N$ algebra and the latter one is from free boson which describes $U(1)$ factor.
While it has limitation to the background charge $Q$, it shows how to integrate
$U(1)$ factor with $W_N$ algebra or Toda fields.

As we mentioned in the introduction, some efforts had been done 
to integrate $U(1)$ factor
with Virasoro current in \cite{Alba:2010qc,Belavin:2011js}
for $N=2$ case.  Let us briefly review some relevant materials in \cite{Alba:2010qc}.

In order to describe the chiral correlators, 
the authors introduce two chiral algebras,
{\em i.e.}~Virasoro algebra described by $L_n$ and $U(1)$ current described by a free boson $a_n$.  
They use additional free boson $c_n$ to describe $L_n$ as
\ba
L_n=\sum_{k\neq 0,n} c_k c_{n-k} + i(2P-nQ)c_n\,,\quad
L_0=\frac{Q^2}{4} -P^2 +2\sum_{k>0}c_{-k}c_k\,,
\ea
where $P$ is the momentum of the boson $c_n$ which describe the vertex operator.
Then they propose to introduce a particular basis 
$|P\rangle_{\vec Y}$ with Young tableaux $\vec{Y}=(Y_1, Y_2)$ 
for the Hilbert space described by $a_n$ and $c_n$ such that
(i) the inner product with vertex operator insertion coincides with
$Z_{\mathrm{bf}}$ in \cite{Alday:2009aq} (the factor of Nekrasov's partition function for a bifundamental field):
\ba
\frac{{}_{\vec Y'} \langle P'| V_\alpha |P\rangle_{\vec Y}}{\langle P'| V_\alpha |P\rangle}
\!&=&\!
\mathcal{F}^{\vec Y'}_{\vec Y}(\alpha|P,P')
\nt\!&=&\!
\prod_{i,j=1,2}\prod_{s\in Y_i} (Q-E_{Y_i, Y'_j}(P_i-P_j'|s) -\alpha)
\prod_{t\in Y_j} (E_{Y'_j, Y_i}(P'_j-P_i|t) -\alpha)\quad
\ea
with $\vec P=(P,-P)$, $\vec P'=(P', -P')$ and 
$
E_{X,Y}(P|s):=P-b A_Y(s)+b^{-1}(L_X(s)+1)
$
where $A(s)$/$L(s)$ is the arm/leg length of a Young tableau,
and  (ii) the inner product of these states is diagonal and equals to
$1/Z_\mathrm{vec}$ in \cite{Alday:2009aq} (the inverse of the factor of Nekrasov's partition function for a vector field):
\ba
{}_{\vec Y'} \langle P| P\rangle_{\vec Y}=N_{\vec Y}
\delta_{\vec Y, \vec Y'}\,,\quad 
N_{\vec Y}=\mathcal{F}^{\vec Y}_{\vec Y}(0|P,P)\,.
\ea
Once one finds such basis, one may decompose any correlator as
\ba
\langle \Phi_1\cdots \Phi_n\rangle
\!&=&\! 
\langle \Phi_1 | \Phi_2 \sum_{\vec Y_1} |\tilde P_1\rangle_{\vec Y_1}
\frac{1}{N_{\vec Y_1} } {}_{\vec Y_1}\langle P_1|\Phi_3\cdots
\Phi_{n-2}\sum_{\vec Y_{n-3}} |\tilde P_{n-3} \rangle_{\vec Y_{n-3}}
\nt&&\!\times
\frac{1}{N_{\vec Y_{n-3}} } {}_{\vec Y_{n-3}}\langle P_{n-3}|\Phi_{n-1}| \Phi_n\rangle
\ea
which coincides with Nekrasov's partition function by construction
after replacing $\Phi_i$ to vertex operators.

In \cite{Alba:2010qc}, the authors gave the explicit form of 
the basis $|P\rangle_{\vec Y}$ when one of $Y_i$ is null ($\emptyset$).  
For such cases, it is 
given as the Jack symmetric polynomial
Jac$_Y(x_1,\cdots, x_{|Y|})$ with its coupling constant $-b^2$ or $-1/b^2$.
Here the power symmetric polynomials of 
arguments $x_1,\cdots, x_{|Y|}$ are given in terms of linear combination
of oscillators
$
\sum_{n} (x_n)^k\propto a_{-k} \pm c_{-k}
$
where $\pm$ depends on which $Y_i$ is null.
For generic $\vec Y$, the explicit construction of the states
$|P\rangle_{\vec Y}$ is difficult 
and the authors gave the algorithm for the construction.

Later in \cite{Belavin:2011js}, Belavin found that the construction
of the basis is simplified when $Q=0$. Namely, the basis can be defined 
by the product of two Schur polynomials
\ba\label{BBs}
|P\rangle_{\vec Y} = s_{Y_1}(x^{(1)})\, s_{Y_2}(x^{(2)} ) 
\ea
where the power symmetric polynomials of $x$ and $y$ are
\ba\label{xy2ac}
x^{(1)}_k\propto a_{-k} + c_{-k}\,,\quad
x^{(2)}_k\propto a_{-k} - c_{-k}\,.
\ea

Now let us compare their construction with ours.
It is well-known that Schur polynomial can be interpreted as
the natural diagonal basis of free fermion system
(see, for example, appendix B in \cite{Awata:1994tf} where
concise review is given).  Therefore, the state (\ref{BBs})
is a basis of two fermion system.  It is natural to compare it
to $N=2$ case in our setup.

In \cite{Alba:2010qc,Belavin:2011js}, the authors did not provide
why particular combinations (\ref{xy2ac}) are needed to construct
basis.  On the other hand, in our approach, 
this exactly corresponds to how Virasoro symmetry is obtained
from $\Winf$ when $U(1)$ factor is separated:
\ba
\phi^{(1)}=\varphi^U+\varphi^V,\quad
\phi^{(2)}=\varphi^U-\varphi^V,
\ea
where $\varphi^U$ gives $a_n$ and $\varphi^V$ gives $c_n$.
In \S\,\ref{sec:chain}, 
we give a detailed study to derive the chain vector
by using free fields.  

While such coincidence might seem to be accidental, one can proceed to
consider $N>2$ case as well.  The next nontrivial case is $N=3$ where
Fock space of $W_3$ algebra generated by $L_{-n}$ and $W_{-n}$.
In our description, the orthogonal basis $\phi^{(i)}$ ($i=1,2,3$) are provided
from free bosons as
\ba\label{redef}
&&\back
\phi^{(1)}=\frac{1}{\sqrt{3}}\tilde\varphi^U+\frac{1}{\sqrt{2}} \varphi^V_1+
\frac{1}{\sqrt{6}} \varphi^V_2
\,,\quad
\phi^{(2)}=\frac{1}{\sqrt{3}}\tilde\varphi^U
-\frac{1}{\sqrt{2}}  \varphi^V_1+\frac{1}{\sqrt{6}}\varphi^V_2\,,
\nt&&\back
\phi^{(3)}=\frac{1}{\sqrt{3}}\tilde\varphi^U-\sqrt{\frac{2}{3}} \varphi^V_2\,,
\ea
where we have changed normalization of free boson for
$U(1)$ part
$\varphi^U \to \frac{1}{\sqrt{3}}\tilde\varphi^U$ 
compared with eq.\,(\ref{n3-varphi}).
In this normalization,
$\tilde\varphi^U(z) \tilde\varphi^U(0)\sim\ln z$
as $\varphi^V_{1,2}$ satisfy.
Therefore, we would like to see such linear combinations give a generalization
of the diagonal basis as
\ba
&&\back
|\vec P\rangle_{\vec Y}\sim s_{Y_1} (x^{(1)})\, s_{Y_2} (x^{(2)})\,
s_{Y_3} (x^{(3)}) 
\ea
where $x^{(1,2,3)}$ are the polynomial representation
of $\phi^{(1,2,3)}$.
In the following sections, we show that the chain vector, once expanded by
this basis, have coefficients which will reproduce Nekrasov's
formula correctly as AGT-W conjecture predicts.


\section{Chain vectors}
\label{sec:chain}

\paragraph{Definition of level $n$ chain vector}
Let
$\mathcal{H}_n$ be the level $n$ states 
generated from highest weight state $|\Delta \rangle$ 
by the action of generators of a chiral algebra.
For example, for $W_3$ algebra, it is generated by $L_{-n}$ and $W_{-n}$
from $|\vec p\rangle$. 

Let $|u_i\rangle$ be a basis of $\mathcal{H}_n$  
($i=1,\cdots,\mbox{dim}\,\mathcal{H}_n$).
We define a projector onto level $n$ states as
\ba
\Pi^{(n)}_\Delta \!&:=&\! \sum_{i,j} |u_i\rangle S^{-1}_{ij} \langle u_j |
\ea
where $S_{ij} =\langle u_i|u_j\rangle $ is Shapovalov matrix.
It satisfies $O_n \Pi^{(N)}_\Delta=\Pi^{(N-n)}_\Delta O_n$
for any element $O_n$ in the chiral algebra.
Then the chain vector at level $n$ is defined as
\ba
|n\rangle_{\Delta, \Delta_1, \Delta_2}
 \!&:=&\! \Pi^{(n)}_\Delta V_{\Delta_1}(1)  |\Delta_2\rangle \label{chain}
\ea
where the expression on the right hand side should be determined by
the conformal Ward identities.

We note that with the chain vector, one can express the
four point function as their inner product:
\ba
\langle \Delta_1| V_{\Delta_2}(z) V_{\Delta_3}(1) |\Delta_4\rangle
=\sum_{\Delta} \sum_{n=0}^\infty z^n {}_{\Delta,\Delta_2,\Delta_1}\!
\langle n| n\rangle_{\Delta,\Delta_3,\Delta_4}\,.
\ea
For the higher correlator, one has to define a generalization of chain vector
as
\ba
\Pi^{(n)}_{\Delta_1} V_{\Delta_2}(1) \Pi^{(m)}_{\Delta_3}
=:O(n,m)_{\Delta_1,\Delta_2,\Delta_3}
\ea
and compute the product
\ba
\sum_{n_1,\cdots,n_r}\sum_{\Delta} \langle n_1|O(n_1,n_2)
\cdots O(n_{r-1},n_r) |n_r\rangle
\ea
where we omit the weight $\Delta$ in the operators/vectors.
Since this kind of correlator corresponds to instanton contribution
of Nekrasov's partition function, the chain vector gives
a building block to prove AGT conjecture.

\subsection{Chain vector for free boson}
In this case, the highest weight state is $|p\rangle$
and chiral algebra is generated by $a_{-n}$.
Since the basis of the oscillator Hilbert space 
$\{a_{-n_1}\cdots a_{-n_r}|p\rangle\}$
are orthogonal, the projector becomes very simple:  for example,
$
\Pi^{(1)}_{p} =a_{-1}| p\rangle\langle  p|a_1
$.

Therefore, the evaluation of eq.\,(\ref{chain}) involves the calculation of
 correlators of the form $\langle p| a_{n_1} \cdots a_{n_s} V_r(1)|q\rangle$,
but they are also very simple: for example,
$
\langle p |a_n^I V_{r}(1)| q\rangle = 
r \langle p| V_{r}(1)|q\rangle
$.
By solving the recursion formula 
$a_n |N\rangle_{p,r,q} =r|N-n\rangle_{p,r,q}$ which can be proved as
\ba
a_n |N\rangle_{p,r,q} = a_n \Pi^{(N)}_p V_r(1)|q\rangle
=  \Pi^{(N-n)}_p a_n V_r(1)|q\rangle= r\Pi^{(N-n)}_p V_r(1)|q\rangle
=r|N-n\rangle_{p,r,q}\,, 
\ea
one may obtain 
a generating function of chain vectors in a closed form:
\ba
\sum_{n=1}^\infty
|n \rangle_{p,r,q} \zeta^n =
e^{ r \sum_{n=1}^\infty \frac{1}{n} a_{-n}^I \zeta^n}| p\rangle 
\ea
from which one may extract $|n\rangle_{p,r,q}$: for example,
\ba
|1\rangle_{p,r,q} = r  a_{-1}| p\rangle,\quad
|2\rangle_{p,r,q}= \frac12 (r a_{-2}
+(r a_{-1})^2) |\vec p\rangle.
\ea
We note that a chain vector for free boson depends only on the momentum 
of $V_r(1)$.  This is the characteristic feature for free boson
which is not shared by chain vector for Virasoro or $W_3$.

\subsection{Virasoro algebra}

The recursion formula for chain vector is
\ba\label{rec:V}
L_k|N\rangle_{\Delta,\Delta_1,\Delta_2}
 = (\Delta +k \Delta_1-\Delta_2 +N-k)|N-k\rangle_{\Delta,\Delta_1,\Delta_2}\,.
\ea
It may be derived by combining $L_k \Pi^{(N)}_{\Delta}=\Pi^{(N-k)}_\Delta L_k$
and a conformal Ward identity
$$\langle u| L_k V_{\Delta_1} |\Delta_2\rangle = (\Delta +N-k -\Delta_2 
+ k\Delta_1)\langle u|V_{\Delta_1} |\Delta_2\rangle$$ which holds for
any  level $N-k$ state $\langle u|$ from $\langle \Delta|$.

The chain vector may be derived in terms of Virasoro operators.
However, in order to do it, we need to invert Shapovalov matrix
which is complicated.  Therefore, instead of doing it,
one may solve it more directly in terms of free boson.
For $c=1$ case, we have
\ba
L_n=\frac12 \sum_{k} :\! a_{n-k} a_k \!:
\ea
where $[a_n, a_m]=n \delta_{n+m,0}$. Then we write
\ba
a_{-n}=n x_n\,, \quad
a_{n} =\partial_{x_n}\quad
(n>0)
\ea
and express the bosonic Fock space as the polynomials of 
variables $x_n$ ($n=1,2,3,\cdots$).
For example, we rewrite $a_{-n_1}\cdots\, a_{-n_r}|p\rangle$
as $n_1x_{n_1} \cdots\, n_r x_{n_r}$.
Using this correspondence,
we denote $\Psi_N(x)$ to represent
the chain vector $|N\rangle_{\Delta,\Delta_1,\Delta_2}$.
We use the vertex operator representation for primary fields with
\ba
\Delta=\frac{p^2}4, \quad \Delta_1=\frac{r^2}4, \quad \Delta_2=\frac{q^2}4,
\ea
which corresponds to $V_\Delta = e^{p \phi/\sqrt{2}}$, 
$V_{\Delta_1} = e^{r \phi/\sqrt{2}}$ and
$V_{\Delta_2} = e^{q \phi/\sqrt{2}}$.
As a result, the recursion relation (\ref{rec:V}) is written
as the differential equation for $\Psi_N$:
\ba
\left(
\sum_{r=1}^\infty r x_r\partial_{x_{k+r}}+\frac{p}{\sqrt{2}}
 \partial_{x_k}+
\frac12 \sum_{s=1}^{k-1} \partial_{x_s}\partial_{x_{k-s}}
\right)\Psi_N= (\Delta +k \Delta_1-\Delta_2 +N-k)\Psi_{N-k}\,.
\ea
Starting from $\Psi_0=1$, one may solve it recursively.
For example,
\ba\label{psi}
\Psi_1 = \frac{\left(p^2-q^2+r^2\right) x_1}{2\sqrt{2} p}\,,\quad
\Psi_2 = \beta_1 x_1^2 + \beta_2 x_2\,,
\ea
where
\ba
\beta_1 \!&=&\! \frac{(p^2-q^2+r^2)^2-4r^2}{16(p^2-1)}\nt
\beta_2 \!&=&\! \frac{3 p^4-2 p^2 \left(q^2-3 r^2+2\right)-(q-r) (q+r)
   \left(q^2-r^2-4\right)}{8 \sqrt{2} p \left(p^2-1\right)}\,.
\ea

The readers may wonder why the chain vector for Virasoro is rather complicated,
compared with that of the free boson. Actually
if one of the ``momentum conservation" conditions
\ba\label{momentum}
p=\epsilon_q q +\epsilon_r r \quad
(\epsilon_q,\epsilon_r=\pm 1)
\ea
is satisfied,
the chain vector is reduced to that of free boson:
\ba
\sum_{n=0}^\infty \Psi_n \zeta^n
~\rightarrow~
\exp\left(-\frac{\epsilon_r r}{\sqrt{2}} \sum_{n=1}^\infty x_n \zeta^n\right)\,.
\ea
If the conservation is violated, we need some screening currents 
to define the correlator. It explains why such simplification does not generally
occur.

\subsection{$W_3$ algebra}
We can derive the chain vector for $W_3$ algebra similarly, namely by combining
$
O_k \Pi^{(N)}_{\bD} =\Pi^{(N-k)}_{\bD} O_k
$
with $O_k=L_k$ or $W_k$ and the Ward identity for $W_3$ algebra.
We use the label $\bD$ to represent the eigenvalues 
$(\Delta,w)$ for the highest weight representation.

The recursion formula for $L_k$ is the same as eq.\,(\ref{rec:V}). 
For the $W_k$ generators, we use 
eqs.\,(38) and (39) of \cite{Kanno:2010kj}
\ba\label{rec:W}
W_k|N\rangle_{\bD,\bD_1, \bD_2} \!&=&\! \left(\frac{k(k+3)}{2} w_1 -w_2\right)
|N-k\rangle_{\bD,\bD_1, \bD_2}\nn\\
&&\!
 +k\Pi^{(N-k)}_\bD (W_{-1} V_{\bD_1})|\bD_2\rangle
 + (\Pi^{(N-k)}_\bD W_0) V_{\bD_1} |\bD_2\rangle\,.
\ea
In order to make it a closed recursion formula, we need to impose
the level 1 null state condition for $V_{\bD_1}$:
\ba
W_{-1} V_{\bD_1}=\frac{3w_1}{2\Delta_1} L_{-1} V_{\bD_1}\,.
\ea
Then the second term of eq.\,(\ref{rec:W}) can be evaluated by Ward identity for Virasoro.
The third term should be left as it is.
To summarize, the recursion formula for $W_k$ is given as
\ba
W_k|N\rangle_{\bD,\bD_1, \bD_2}
 \!&=&\! \left(\frac{k(k+3)}{2} w_1-w_2+\frac{3kw_1}{2\Delta_1}
(N-k+\Delta-\Delta_1-\Delta_2)\right)|N-k\rangle_{\bD,\bD_1, \bD_2}\nn\\
&&\!+W_0 |N-k\rangle_{\bD,\bD_1, \bD_2}\,.
\ea

Again, we would like to solve these recursion formulae by free boson representation.
If we write
\ba
\partial \phi^1 \!&=&\! p_1 z^{-1} +\sum_{k=n}^\infty x_n z^{n-1}
+ \sum_{n=1}^\infty \frac{z^{-n-1}}n \frac{\partial}{\partial x_n}\,,\nt
\partial \phi^2 \!&=&\! p_2 z^{-1} +\sum_{k=n}^\infty y_n z^{n-1}
+ \sum_{n=1}^\infty \frac{z^{-n-1}}n \frac{\partial}{\partial y_n}\,,
\ea
the oscillator representation for generators $L_k, W_k$ ($k\geq 0$) becomes
{\small
\ba
L_k \!&=&\!
\sum_{r=1}^\infty r x_r\partial_{x_{k+r}}+p_1 \partial_{x_k}+
\frac12 \sum_{s=1}^{k-1} \partial_{x_s}\partial_{x_{k-s}}
+\sum_{r=1}^\infty r y_r\partial_{y_{k+r}}+p_2 \partial_{y_k}+
\frac12 \sum_{s=1}^{k-1} \partial_{y_s}\partial_{y_{k-s}}\\
6 W_k \!&=&\! \sum_{n,m=1}^{n+m< k}  \left(
\frac{\partial^3}{\partial y_n \partial y_m \partial y_{k-n-m}}
-3 \frac{\partial^3}{\partial x_n \partial x_m \partial y_{k-n-m}}
\right)\nn\\
&&\!+3\left(\sum_{\stackrel{\scriptstyle n,m=1}{\scriptstyle n+m>k}}(n+m-k)\left(
y_{n+m-k}\frac{\partial^2}{\partial y_n \partial y_m}
-2 x_{n+m-k}\frac{\partial^2}{\partial x_n \partial y_m}
-y_{n+m-k}\frac{\partial^2}{\partial x_n \partial x_m}
\right)
\right)\nn\\
&&\!+3\left(
\sum_{n,m=1} nm \left( y_n y_m \frac{\partial}{\partial y_{n+m+k}} 
-2x_n y_m \frac{\partial}{\partial x_{n+m+k}}
-x_n x_m \frac{\partial}{\partial y_{n+m+k}}
\right)\right)
\nn\\
&&\!
-3 \left(
 \sum_{n=1}^{k-1}\left( -p_2 \frac{\partial^2}{\partial y_n\partial y_{k-n}}
+ p_2\frac{\partial^2}{\partial x_n\partial x_{k-n}}
+ p_1\frac{\partial^2}{\partial x_n\partial y_{k-n}}
 \right)
\right)\nn\\
&&\! +6 \sum_{n=1}^{\infty} n\left(
p_2 y_n\frac{\partial}{\partial y_{n+k}}
- p_2 x_n\frac{\partial}{\partial x_{n+k}}
- p_1 x_n\frac{\partial}{\partial y_{n+k}}
- p_1 y_n\frac{\partial}{\partial x_{n+k}}
\right)\nn\\
&&\!+3\left((p_2^2-p_1^2) \frac{\partial}{\partial y_{k}}
-2 p_1 p_2  \frac{\partial}{\partial x_{k}}
\right) (1-\delta_{k0})
+(p_2^3-3p_1^1p_2)\delta_{k0}\,.
\ea
}
For level 1, the recursion formula is
\ba
L_1 |1\rangle \!&=&\! (\Delta+\Delta_1-\Delta_2) |0\rangle\nt
W_1 |1\rangle \!&=&\! (2w_1-w_2+w +\frac{3w_1}{2\Delta_1}(\Delta-\Delta_1-\Delta_2)|0\rangle\,.
\ea
If we write $|1\rangle = \alpha_1 x_1+\alpha_2 y_1$,
we obtain
\ba
L_1|1\rangle = \alpha_1 p_1+\alpha_2 p_2\,,\quad
2 W_1|1\rangle = (p_2^2-p_1^2) \alpha_2-2p_1 p_2 \alpha_1\,.
\ea
We assign the momentum $(0,m)$ for $V_1(1)$ and 
$(q_1,q_2)$ for $|V_0\rangle$.  We note that $V_1$ must have a level 1
null state and the assignment for $V_1$ is one possibility for it.

By comparing these formula, one can determine $\alpha_{1,2}$:
\ba
\alpha_1=\frac{A_1}{6\left({p_1}^2-3 {p_2}^2\right)}\,,\quad
\alpha_2=-\frac{A_2}{6\left({p_1}^2-3 {p_2}^2\right)}\,,
\ea
where
\ba
A_1\!&:=&\! 3 {p_1}^4-3 \left(2
   {p_2}^2+{q_1}^2+{q_2}^2\right) {p_1}^2+m^3
   {p_2}+3 m^2 \left({p_1}^2-{p_2}^2\right)\nn\\
&&\!+3 m
   {p_2}
   \left({p_1}^2+{p_2}^2-{q_1}^2-{q_2}^2\right)
-{p_2} ({p_2}+2 {q_2}) \left({p_2}^2-2
   {q_2} {p_2}-3 {q_1}^2+{q_2}^2\right)\\
A_2\!&:=&\! m^3+6 {p_2} m^2+3
   \left({p_1}^2+{p_2}^2-{q_1}^2-{q_2}^2\right) m+2 ({p_2}-{q_2}) \left(4 {p_2}^2+4
   {q_2} {p_2}-3 {q_1}^2+{q_2}^2\right).\nonumber
\ea
The denominator factor $p_1^3-3p_1 p_2^2$ vanishes when
$p_1=0$ or $p_1=\pm \sqrt{3} p_2$. This is precisely
the correct momentum to have level 1 null state.
Another consistency check is that it reduces to the free boson
chain vector once we impose the momentum conservation law
\ba
p_1=q_1,\quad p_2=m+q_2
\quad\rightarrow\quad
\alpha_1=0,\quad \alpha_2=m\,.
\ea

\section{Combination with $U(1)$ part: comparison with gauge theory}
\label{s:comb}
The claim in \cite{Belavin:2011js} is that 
once the chain vector is combined with $U(1)$ part 
and re-expanded in terms of Schur polynomial, its coefficients
of expansion implies Nekrasov's formula.
  
The chain vector for $U(1)$ part is written in the form
\ba
\Psi^U_p(\zeta) = \sum_{n=0}^\infty \Psi_{p,n}^{U} \zeta^n =
e^{p \sum_{s=1}^\infty t_n \zeta^n}.
\ea
We mix it with chain vector as
\ba\label{Ptot}
\Psi(\zeta) =\Psi^U(\zeta) \Psi^V(\zeta)
\,,\quad\text{or}\quad
\Psi_N= \sum_{n=0}^N \Psi^U_n \Psi^V_{N-n}
\ea 
where $\Psi^V(\zeta)$ is the generating function for Virasoro or $W_3$ algebra.
Let us first reproduce the results of \cite{Belavin:2011js} for
Virasoro case.

\subsection{Virasoro vs. $SU(2)$ gauge theory}

We have already given the explicit form of chain vector
for Virasoro algebra in eq.\,(\ref{psi}).
We computed the result up to level 3 but do not write it here,
since it is complicated and not illuminating.

For $\Psi_{\Delta,\Delta_1,\Delta_2}$ with
$\Delta=p^2/4$, $\Delta_1=r^2/4$ and $\Delta_2=q^2/4$,
we choose $U(1)$ part to be $\Psi^U_{r/\sqrt{2}}$.
In $\Winf$ representation, it implies that we need use the representation
$(\lambda_1,\lambda_2)=(r,0)$ in eq.\,(\ref{sDE}) for $C_i=1$ and $K=2$.
After the combination as eq.\,(\ref{Ptot}) and the change of variables as
\ba
x_n=(\tilde x_n-\tilde y_n)/\sqrt{2}\,,\quad
t_n=(\tilde x_n+\tilde y_n)/\sqrt{2}\,,
\ea
we get an expansion of the form 
\ba
\Psi_N = \sum_{|Y_1|+|Y_2|=N} C(Y_1, Y_2)\, s_{Y_1}(\tilde x)\, s_{Y_2}(\tilde y)
\ea
where $s_Y(x)$ is the Schur polynomial in terms
of power sum polynomial.  For example, up to level 3,
\ba
&&\back
s_{\emp}(x)=1\,,\quad
s_{[1]}(x)=x_1\,,\quad
s_{[2]}(x)=\frac{x_1^2}{2}+x_2\,,\quad
s_{[1^2]}(x) =\frac{x_1^2}{2}-x_2\,,\nn\\
&&\back
s_{[3]}(x)=\frac{x_1^3}{6}+x_1 x_2 +x_3\,,\quad 
s_{[1^3]}(x)=\frac{x_1^3}{6}-x_1 x_2 +x_3\,,\quad
s_{[2,1]}(x)= \frac{x_1^3}{3}-x_3\,.
\ea
The coefficient $C(Y_1, Y_2)$ is written of the form
\ba\label{coeff:V}
C(Y_1, Y_2) =\frac{z(Y_1,p+q+r)\,z(Y_1,p-q+r)\,z(Y_2^t,p+q-r)\,z(Y_2^t,p-q-r)}{
D(\vec Y,p)}
\ea
where 
\ba
z(Y,x)&=&\prod_{(k,l)\in Y} (x/2+k-l)\nt
D(\vec Y,p)&=& \prod_{i,j=1}^2\prod_{s\in Y_i}\bigl(a_i-a_j+A_i(s)+L_j(s)+1\bigr) 
\ea
and $a_1=p/2$, $a_2=-p/2$. 
$s=(k,l)$ denotes the position of the box in a Young tableau 
({\em i.e.}~the box in $k$-th column and $l$-th row).
$A(s)$/$L(s)$ is the arm/leg length of a Young tableau, respectively. 
In particular, for the level $N=1,2,3$,
\ba
&&\back
D(([1],\emp),p)=p\,,\quad 
D((\emp,[1]),p)=-p\,,\nn\\
&&\back
D(([2],\emp),p)=2p(p+1)\,,\quad
D(([1^2],\emp),p)=2p(p-1)\,,\nt
&&\back
D(([1],[1]),p)=-(p+1)(p-1)\,,\nn\\
&&\back
D((\emp,[2]),p)=2p(p-1)\,,\quad
D((\emp,[1^2]),p)=2p(p+1)\,,\nn\\
&&\back
D(([3],\emp),p)= 6p(p+1)(p+2)\,,\quad 
D(([2,1],\emp),p)= 3p(p+1)(p-1)\,,\quad \nn\\
&&\back
D(([1^3],\emp),p)= 6p(p-1)(p-2)\,,\nn\\
&&\back
D(([2],[1]),p)= -2 p(p-1)(p+2)\,,\quad  
D(([1^2],[1]),p)= -2 p(p+1)(p-2)\,,\nt
&&\qquad\vdots
\ea
Therefore, we can confirm that 
the coefficients (\ref{coeff:V}) exactly correspond to Nekrasov's partition function with $\eps_1/\eps_2=-1$.

\subsection{$W_3$ vs. $SU(3)$ gauge theory}

We note that, as in the Virasoro case, the chain vector is
constructed out of the free boson $\tilde\varphi^V_{1,2}$.
Now we need to combine it with $U(1)$ part $\tilde\varphi^U$ and
rewrite the combined chain vector in terms of $\phi^{(i)}$.
The relation between them is given in eq.\,(\ref{redef}) as
\ba
&&\back
\tilde\varphi^V_1=\frac{1}{\sqrt{2}} (\phi^{(1)}-\phi^{(2)})\,,\quad
\tilde\varphi^V_2=\frac{1}{\sqrt{6}} (\phi^{(1)}+\phi^{(2)}-2\phi^{(3)})\,,\nt
&&\back
\tilde\varphi^U=\frac{1}{\sqrt{3}} (\phi^{(1)}+\phi^{(2)}+\phi^{(3)})\,.
\ea
We also rewrite the momentum by those for the orthogonal basis:
\ba
&&\back
p_1=\frac{1}{\sqrt{2}}(a_1-a_2)\,,\quad
p_2=\frac{1}{\sqrt{6}}(a_1+a_2-2 a_3)\,,\quad
p_3=\frac{1}{\sqrt{3}}(a_1+a_2+a_3)\,,\nt
&&\back
q_1=\frac{1}{\sqrt{2}}(b_1-b_2)\,,\quad
q_2=\frac{1}{\sqrt{6}}(b_1+b_2-2 b_3)\,,\quad
q_3=\frac{1}{\sqrt{3}}(b_1+b_2+b_3)\,,\nt
&&\back
r_1=\frac{1}{\sqrt{2}}(c_1-c_2)=0\,,\quad
r_2=\frac{1}{\sqrt{6}}(c_1+c_2-2 c_3)=m\,,\quad
r_3=\frac{1}{\sqrt{3}}(c_1+c_2+c_3)\,.
\quad
\ea
where $p_3,q_3,r_3$ are momenta for the $U(1)$ factor.
$a_i,b_i, c_i$ are momenta for orthogonal basis $\phi^{(i)}$.
We need to impose $r_1=0$ for the corresponding vertex to
a level 1 null state which is necessary to solve
conformal Ward identity.

The chain vector is written as
\ba
\psi_1=\psi_1^V+\psi_1^U\,;\quad
\psi_1^V=\alpha_1 x_1+\alpha_2 x_2\,,\quad
\psi_1^U= r_3 t\,.
\ea
where $t$ is the variable for $U(1)$ boson.
Then we need to use the following assignment to proceed:
\ba
c_1=c_2=0\,,\quad
c_3=3c\,.
\ea
From the viewpoint of $\Winf$ representation, this assignment is
equivalent to impose $\lambda_1=\lambda_2=0$ in eq.\,(\ref{sDE2})
while leaving $\lambda_3$ arbitrary.
We note that such assignment was also used for $SU(2)$ case.
We guess that similar assignment will be necessary also
for higher cases $N>3$.

Another comment is that we also need to impose $p_3=q_3=0$
in the $W_3$ chain vector to give the correct formula.
This is natural since these parameters are momenta for $U(1)$
which is irrelevant in the representation of $W_3$ algebra.

We need to rewrite the oscillator similarly:
\ba
x_1=\frac{1}{\sqrt{2}}(X_1-X_2)\,,\quad
x_2=\frac{1}{\sqrt{6}}(X_1+X_2-2 X_3)\,,\quad
t=\frac{1}{\sqrt{3}}(X_1+X_2+X_3)\,.
\ea
In terms of these basis, the level 1 chain vector has following
factorized form
\ba
\psi_1=\gamma_1 X_1+\gamma_2 X_2+\gamma_3 X_3
\ea
with
\ba
\gamma_1\!&=&\! \frac{ ( a_1- b_1+c) (a_1-b_2+c) ( a_1- b_3+c)}{ (a_1-a_2) (
   a_1-a_3)}\nt
\gamma_2 \!&=&\!
\frac{( a_2- b_1+c) (a_2-
   b_2+c) ( a_2- b_3
   +c)}{ ({a_2}-{a_1})
   ( {a_2}-a_3)}
\nt
\gamma_3 \!&=&\!
\frac{( a_3-b_1+c) (
   {a_3}-b_2+{c})
   (a_3- b_3+c)}{ (
   a_3-a_1) (a_3- a_2)}
\ea
This again takes the expected form, {\em i.e.}~the denominator 
factor corresponds to the factor of Nekrasov's partition function for a vector field
and the numerator takes the form of that for fundamental or 
anti-fundamental matter fields.

We conclude that $\Winf$ symmetry seems to play a critical
role in how to recombine of free fields.
It also seems to be essential in choosing the momentum for 
the intermediate vertex operator in the form $(\lambda,0,\cdots,0)$.
Also, the denominator factor vanishes when the weight $\Delta$
has the form $\lambda_i-\lambda_j=$ integer
which is exactly the null state condition suggested from eq.\,(\ref{bn}).

\section{Conclusion}
\label{s:conclusion}

In this paper, we argue that $\Winf$ algebra explains
the correct inclusion of $U(1)$ factor to the symmetry
of Toda fields.  It also gives any $W_N$ symmetry in the same footing,
namely it reduces to choosing correct 
quasi-finite unitary representations.
In this sense, it should be regarded as the correct symmetry
behind AGT-W relation. 

The reader may have some criticism on our identification
of U(1) factor is merely the enhancement of $SU(N)$ to
$U(N)$.  We would like to argue, however, that $\Winf$ algebra
automatically contains infinite commuting charges $W(D^n)$
which would be helpful to understand exactly solvable
system behind such correspondence.

Of course, the computation made here still depends heavily on the original
$W$ symmetry.  In this sense, we have not utilize the full machinery of
the symmetry.  For example, in the computation of chain vector
made in \S\,\ref{sec:chain}-\ref{s:comb},
we can not use $\Winf$ algebra directly.
A direct proof in \cite{Mironov:2010pi}, where Selberg integral is 
performed, might be helpful.

Since $\Winf$ has much simpler structure than $W_N$ algebra, 
it is easy to convince ourselves that factorization of 
Nekrasov's formula may directly come from $\Winf$ symmetry.
So far, we have not achieved it since we do not know how to define
the three point functions which seemed not be studied in the literature.

Such computation would be also useful to give us some inspiration
to understand non-Lagrangian strong coupling theories which
was conjectured by Gaiotto \cite{Gaiotto:2009we}.
In case of $W_N$, it was difficult to calculate corresponding
correlation function since conformal Ward identity could
not be solved.  For $\Winf$ case, however, it has much higher symmetry
and one may have some hope to define the correlator.

Another material which we can not study so far is the general case
$Q\neq 0$.  Since $\Winf$ algebra is limited to describe $C=N$, we need
some sort of deformation.  Judging from the observation in \cite{Alba:2010qc},
it will be natural consider the interacting system (Calogero-Sutherland),
to guess the symmetry behind it.  A generalization of exactly solvable
system in the appendix of \cite{Alba:2010qc} would be promising direction.
We note that Jack polynomial has an interpretation of null states
of $W_N$ algebra \cite{b:Jack-W}. See also a work \cite{Kac:1996nq}
where general CFT was studied in the context of $\Winf$ algebra.

For the extension AGT conjecture to $SU(N)$ linear
quiver gauge theory,
we have conjectured that general level 1 null state
describes the general puncture \cite{Gaiotto:2009we,Kanno:2009ga}.
This correspondence seems to have some subtleties
as found later \cite{Kanno:2010kj}.  In \cite{Drukker:2010vg}, authors
gave a  proposal which would be possible solution to the problem.
However, the $U(1)$ seems to be involved if we examine
level higher than 2.
We hope that $\Winf$ symmetry
provide some hints to this issue.

Moreover, the discussion on surface operator in 
$SU(N)$ gauge theory is also an interesting topic.
For $SU(2)$ case, it is already known that the 
corresponding operator in Liouville theory is related 
to level 2 null states \cite{Alday:2009fs,b:sr}. 
Then it is natural to expect that the corresponding operator 
in Toda theory is related to higher level null states. 
It is very complicated to classify them in $W_N$ algebra, 
but from the viewpoint of $W_{1+\infty}$ 
algebra, this discussion may become much simpler.

$\Winf$ symmetry has been applied to many topics,
for example,  the quantum Hall effect
\cite{Cappelli:1995ts}, matrix model \cite{b:matrix}
(see also a recent development in the context of AGT
\cite{b:mn}), 
topological string \cite{Dijkgraaf:1996iy} and crystal melting
\cite{Heckman:2006sk}. We hope that it is a good time
now to develop the representation theory, such as the correlation
function, to more detail.

\paragraph{Acknowledgements}
S. K. is partially supported by Grant-in-Aid 
(\#23-10372) for JSPS Fellows.
Y. M. is partially supported by Grant-in-Aid (\#20540253) 
from MEXT, Japan.
S. S. is partially supported by Grant-in-Aid 
(\#23-7749) for JSPS Fellows.



\appendix
\section{Summary of convention}
Free fields and vertex
\ba
&&\back
\phi^I(z) \phi^J(w) \sim \delta^{IJ} \log (z-w)\,,\quad
\partial_z \phi^I= \sum_{n} a_n^I z^{-n-1}\,,\nt
&&\back
\left[ a_n^I, a_m^J \right] = n \delta_{n+m,0}\, \delta^{IJ}\,,\quad
a^I_0\equiv  \hat p^I\,,\nt
&&\back
\partial\phi^I(z)V_{\vec p}(0) \sim \frac{p^I}{z}V_{\vec p}(0)\,,\quad 
\lim_{z\rightarrow 0}V_{\vec p}(z) |0\rangle = |\vec p\rangle\,,\quad
 \hat p |\lambda\rangle = \lambda|\lambda \rangle\,.
\ea 
Noether currents of Virasoro and $W_3$
\ba
T(z) = \frac12 :\!(\partial\phi^1)^2\!:+\frac12 :\!(\partial\phi^2)^2\!:\,,\quad
W(z) = \frac16 \bigl(:\!(\partial \phi^2 )^3\!: -3:\!(\partial\phi^1 )^2\partial\phi^2\!: \bigr)
\ea
Conformal and $W_3$ weight
\ba
&&\back
L_0|\vec p\rangle = \Delta(\vec p)|\vec p\rangle\,,\quad
W_0|\vec p\rangle = w(\vec p)|\vec p\rangle\,,\nn\\
&&\back
\Delta(\vec p)=\frac12 ((p^1)^2+(p^2)^2)\,,\quad
w(\vec p)=\frac{1}{6}((p^2)^3-3 p^2(p^1)^2)
\ea

\end{document}